\documentclass{svproc}
\usepackage{url}

\usepackage[
   colorlinks,
   linkcolor=blue,
   urlcolor=blue,
   citecolor=blue,
   ]{hyperref}
\usepackage{multicol}
\usepackage{graphicx}
\usepackage{setspace}
\usepackage{enumerate}
\usepackage{cite}
\usepackage{amsmath}
\usepackage[hang,flushmargin]{footmisc} 

\begin{document}
\mainmatter              
\title{Attosecond Physics and Quantum Information Science}
\titlerunning{Attosecond Physics and Quantum Information Science}  
%
\author{
M.~Lewenstein\inst{1,2} \and N.~Baldelli\inst{1} \and U.~Bhattacharya\inst{1} \and J.~Biegert\inst{1,2} \and M.F.~Ciappina\inst{3,4} \and U.~Elu\inst{1} \and T.~Grass\inst{1} \and P.T.~Grochowski\inst{1,5,6,7} \and A.~Johnson\inst{1} \and Th.~Lamprou\inst{8,12} \and A.S.~Maxwell\inst{9} \and A.~Ordóñez\inst{1} \and E.~Pisanty\inst{10,11} \and J.~Rivera-Dean\inst{1} \and P.~Stammer\inst{1} \and I.~Tyulnev\inst{2} \and P.~Tzallas\inst{12,13}
}

%
\authorrunning{Maciej Lewenstein et al.} 
%
\tocauthor{Maciej Lewenstein, et.}
\institute{ICFO - Institut de Ciencies Fotoniques, The Barcelona Institute of Science and Technology, 08860 Castelldefels (Barcelona), Spain\\
\and
ICREA, Pg.~Lluís Companys 23, 08010 Barcelona, Spain\\
\and
Physics Program, Guangdong Technion - Israel Institute of Technology, Shantou, Guangdong
515063, China
\and
Technion - Israel Institute of Technology, Haifa, 32000, Israel
\and
Institute for Quantum Optics and Quantum Information of the Austrian Academy of Sciences, A-6020 Innsbruck, Austria
\and
Institute for Theoretical Physics, University of Innsbruck, A-6020 Innsbruck, Austria
\and
Center for Theoretical Physics, Polish Academy of Sciences, Aleja Lotnik\'ow 32/46, 02-668 Warsaw, Poland
\and
Department of Physics, University of Crete, P.O. Box 2208, GR-70013 Heraklion (Crete), Greece
\and
Department of Physics and Astronomy, Aarhus University, DK-8000 Aarhus C, Denmark
\and
Max Born Institute for Nonlinear Optics and Short Pulse Spectroscopy, Max Born Strasse 2a,
D-12489 Berlin, Germany
\and
Department of Physics, King’s College London, Strand, London, UK
\and
Foundation for Research and Technology-Hellas, Institute of Electronic Structure \& Laser, GR-
70013 Heraklion (Crete), Greece
\and
ELI-ALPS, ELI-Hu Non-Profit Ltd., Dugonics tér 13, H-6720 Szeged, Hungary\\
\email{maciej.lewenstein@ifco.eu}}

\maketitle              

\begin{abstract}
In this article,  we will discuss a possibility of 
a symbiosis for attophysics (AP) and quantum
information (QI) and quantum technologies (QT). We will argue that within few years AP will reach Technology Readiness Level (RTL) 4-5 in QT, and will thus become a legitimate platform for QI and QT.
\end{abstract}
\section{Introduction}

Contemporary Quantum Technologies face major difficulties in fault tolerant quantum computing with error correction, and focus instead on various shades of quantum simulation (Noisy Intermediate Scale Quantum, NISQ) devices \cite{Preskill18}, analogue and digital Quantum Simulators \cite{GAN14}, and quantum annealers \cite{FFG01}. There is a clear need and quest for systems that, without necessarily simulating quantum dynamics of some physical systems, can generate massive, controllable, robust entangled and superposition states. This will, in particular, allow for the use of decoherence in a controlled manner, enabling the use of these states for quantum communications \cite{GT07} (e.g. to achieve efficient transfer of information in a safer and quicker way), quantum metrology \cite{GLM11}, sensing and diagnostics \cite{DRC17} (e.g. to precisely measure phase shifts of light fields, or to diagnose quantum materials).

In this lecture, we propose an answer to these needs, by opening new avenues for Quantum Information (QI) science via the symbiosis with Attophysics (AP) and Quantum Optics (QO). To date, there are no existing platforms that can bring processes at such short time-scales to Quantum Information systems. We will illustrate how recent developments in AP aim at realizing a universal and firmly established tool to offer completely unknown solutions and avenues for QI. In particular, we will discuss how AP offers a set of stable and reproducible methods to generate massively entangled states and massive quantum superpositions \cite{RLP22,SRL22}. These methods apply, in the first place, to fundamental Quantum Information science, but with the final goal of bringing them to Quantum Technologies (QT). This will be accomplished by studying:
\begin{itemize}
\item[i)]{The detection of topology, strongly correlated systems, chirality, etc. in AP;}
\item[ii)]{The generation of entangled/quantum correlated states using conditioning methods;}
\item[iii)]{Strong-field physics and atto-second science driven by non-classical light;}
\item[iv)]{The generation of entangled/quantum correlated states in {\it Zerfall} processes;}
\item[v)]{Quantitative and measurable effects of decoherence in AP.}
\end{itemize}

The lecture is organized around four projects that are being realized at ICFO: 
\begin{itemize}
\item European Research Council Advanced Grant "NOvel Quantum SImulAtors (NOQIA)", run by M. Lewenstein. 
\item European Research Council Advanced Grant "Structural transformations and phase transitions in real-time (TRANSFORMER)", run by J. Biegert. 
\item European Future and Emerging Technologies Project "Optical Topological Logic" (OPTOLOGIC), coordinated by J. Biegert (ICFO). 
\item Not-yet-financed project Attophysics and Quantum information Science (ATTOQUIS), coordinated by M. Lewenstein.
This project summons experts in the fields of QO, QI, and AP who will join forces to invent new mechanisms for generating massive, controllable, robust, entangled light states, and understand the process of decoherence in them, thus identifying new laws and limits of QI. In a nutshell:  Our results will set up the basis for a road map toward a novel platform of AS for QT.
\end{itemize}

\begin{figure}[t]
    \centering
    \includegraphics[width=\textwidth]{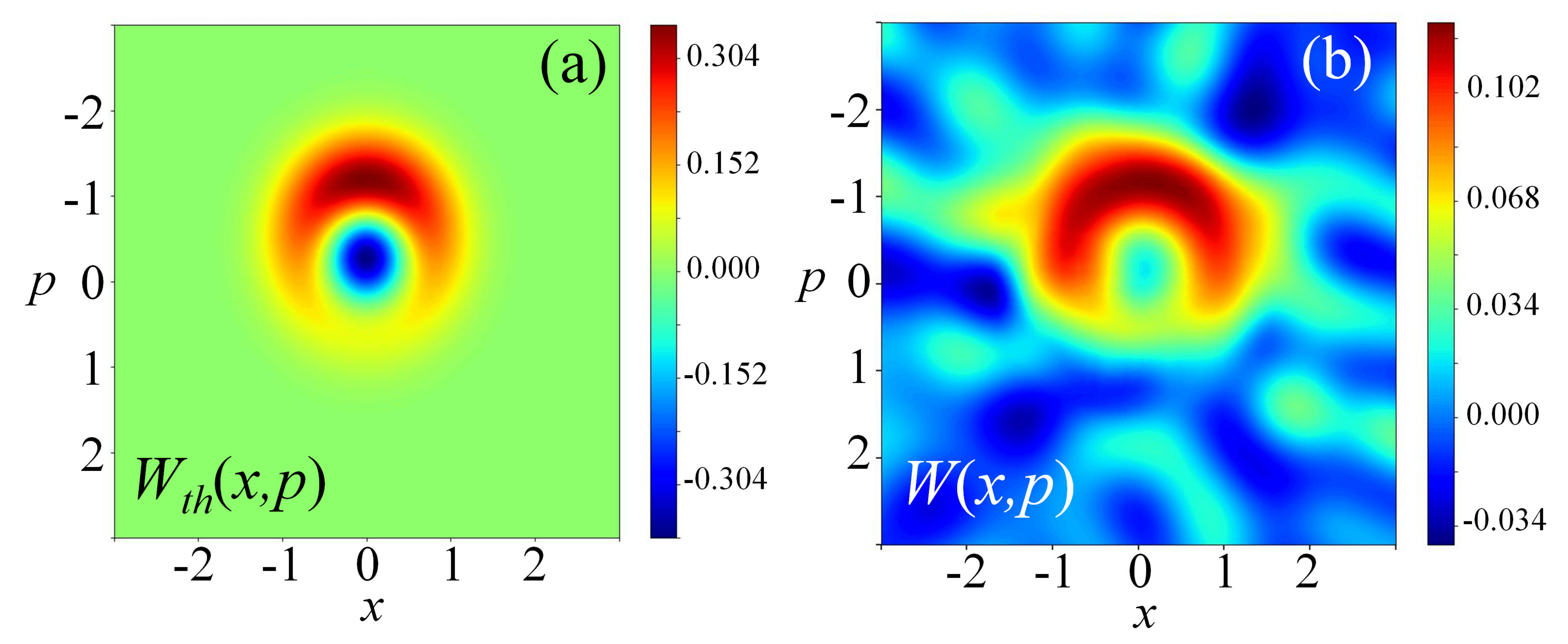}
    \caption{High-photon number optical cat state created by conditioning on HHG. (a) Theoretical prediction for the Wigner function according to \cite{LCP21,RLP22}. (b) Experimental reconstruction of the Wigner function with mean photon number $\langle n \rangle \approx 9.4 \pm 0.1$, which represents an optical Schrödinger cat state. For visualization reasons, the plots have been centered to zero. The figure has been extracted from \cite{RLP22}.}
    \label{Fig:Cats}
\end{figure}

In this lecture we will focus on ATTOQUIS (to be read as ATTOKISS), the new joint project of ICFO, FORTH, Technion, CEA, and IOTA toward a symbiosis of attophysics and quantum information science. This is in particular aimed at studying the generation of entangled/quantum correlated states using conditioning methods. We will describe our joint efforts to generate Schrödinger cat states (SCS) of photons conditioned on high harmonic generation (HHG) \cite{LCP21} (see Fig.~\ref{Fig:Cats}) and/or above threshold ionization \cite{RLP22,SRM22}. We will also talk about generation of topological order using laser pulses with orbital angular momentum (OAM) \cite{BCG22}. We will also discuss HHG as a tool to detect phase transitions and topological order in strongly correlated systems \cite{ABB22,BBG22}. Finally, we will speculate about measurement of entanglement of OAM of electrons in double ionization \cite{MML21}, and violation of Bell inequalities by SCS. This paper is by no means a review -- it is rather a review of progress, based on a selection of relevant references, actually mostly from external and collaborating groups. It follows the structure of the invited lecture of M. Lewenstein at ATTOVIII Conference in Orlando \cite{ATTOVIII}.

\section{Detection of topology, strongly correlated systems, chirality, etc.}

Recently, there is an explosion of interest in applying AP methods to diagnose and detect topological order, strongly correlated systems, and chirality. The primary role is played here by the process of high harmonic generation (HHG). 

\subsection{Detection of topological order with HHG}

The pioneering work of R. Huber {\it et consortes} \cite{RSS18} studied lightwave-driven Dirac currents in a topological surface band. It was noted that ``Harnessing the carrier wave of light as an alternating-current bias may enable electronics at optical clock rates'' \cite{RSS18,KS14}. ``Lightwave-driven currents have been assumed to be essential for high-harmonic generation in solids'' (cf. \cite{HLS15,GR19}). They report angle-resolved photoemission spectroscopy with subcycle time resolution that enables them to observe directly how the carrier wave of a terahertz light pulse accelerates Dirac fermions in the band structure of the topological surface state of Bi$_2$Te$_3$. This work opened a path toward  all-coherent lightwave-driven electronic devices. 

\begin{figure}[t]
    \centering
    \includegraphics[width=\textwidth]{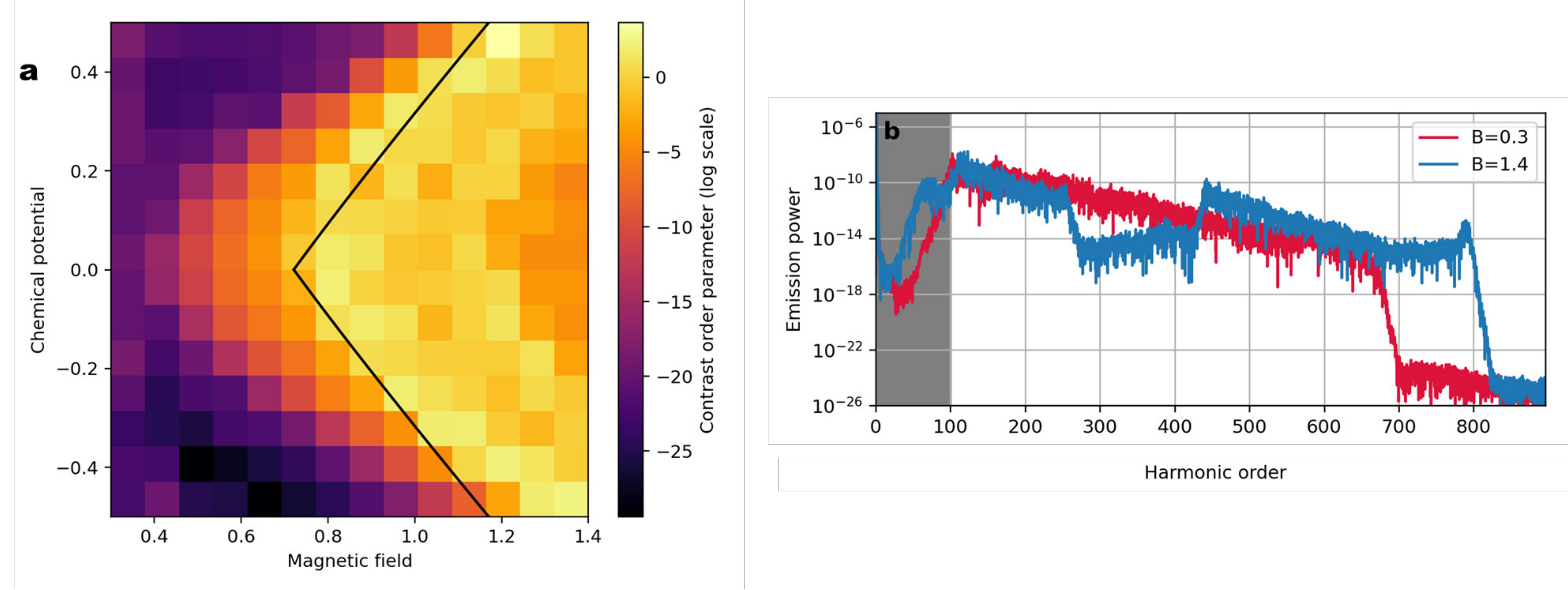}
    \caption{ (a) The phase diagram of the proximity-induced p-wave superconductor is obtained from an order parameter based on the harmonic emission spectrum, exemplarily shown in (b) for two parameter choices, in the topological and in the trival phase. At frequencies below the bandgap, emission is suppressed in the trivial phase (red), whereas in the topological phase emission is possible already at frequencies above half the bandgap due to the edge state in the middle of the bandgap. The figures have been reprinted from Ref. \cite{BBG22}.}
    \label{Fig:p-wave}
\end{figure}

Perhaps the first papers where HHG was considered for detection of topology were the series of papers by D. Bauer with his collaborators on detection of topological edge states in dimerized chains \cite{BH18,JB19}.
In the first work, HHG in the two topological phases of a finite, one-dimensional, periodic structure is investigated using a self-consistent time-dependent density functional theory approach. For harmonic photon energies smaller than the band gap, the harmonic yield is found to differ by up to 14 orders of magnitude for the two topological phases. Similar conclusions were obtained in the study of the  Su-Schrieffer-Heeger (SSH) chains  that display topological edge states. The authors calculated high-harmonic spectra of SSH chains that are coupled to an external laser field of a frequency much smaller than the band gap. In recent works, this method  has been extended to the detection of Majorana fermions in the Kitaev chain \cite{pattanayak21} and in quantum wires with proximity-induce p-wave superconductivity \cite{BBG22} (see Fig.~\ref{Fig:p-wave}). Specifically, their harmonic emission spectrum in strong fields is shown to exhibit spectral features due to radiating edge modes, which characterize the spectrum and the density of states in the topological phase, and which are absent in the trivial phase. These features allow us to define an order parameter, obtained from emission measurements, that unambiguously differentiates between the two phases. Local probing provides insight into the localized and topologically protected nature of the modes. The presented results establish that high harmonic spectroscopy can be used as a novel all-optical tool for the detection of Majorana zero modes.

In the seminal Nature Photonics paper, M. Ivanov {\it et consortes} discuss topological strong-field physics on sub-laser-cycle timescale \cite{SJA19}. These authors study response of the paradigmatic Haldane model \cite{Hal88} to ultrashort, ultrastrong laser pulses. Here, they show that electrons tunnel differently in trivial and topological insulators, for the same band structure, and identify the key role of the Berry curvature in this process. These effects map onto topologically dependent attosecond delays and helicities of emitted harmonics that record the phase diagram of the system. As they say: ``These  findings create new roadmaps in studies of topological systems, building on the ubiquitous properties of the sub-laser-cycle strong-field response—a unique mark of attosecond science''. Moreover, they show that strong fields can also be used for the manipulation of topological properties of 2D materials, relevant for valleytronic devices \cite{JSS20}. In a complementary approach, A. Chacón {\it et al.} \cite{CZK20} studied circular dichroism in higher-order harmonic generation from the Haldane model. These effect clearly heralds topological phases and transitions in  Chern insulators (see also A. Chacón's contribution to \cite{ATTOVIII} and \cite{Heide2022} for very recent findings).\\

\subsection{Detection of strongly correlated systems with HHG}

This is another rapidly developing area. As pointed out in  the recent review \cite{GB22}: ``A hallmark effect of extreme nonlinear optics is the emission of high-order harmonics of the laser from the bulk of materials. The discovery and detailed study of this phenomenon over the course of the past decade has offered a broad range of possibilities and seen the dawn of a new field of extreme solid-state photonics. In particular, it opens the way to previously inaccessible spectral ranges, as well as the development of novel solid-state spectroscopy and microscopy techniques that enable detailed probing of the electronic structure of solids.''

An example of pioneering studies of HHG in strongly correlated materials is described in the paper by Silva {\it et al.} \cite{SBA18}, where the authors show that HHG  can be used to time-resolve ultrafast many-body dynamics associated with an optically driven phase transition to Mott insulator state, with accuracy far exceeding one cycle of the driving light field. In Ref. \cite{BHL21}, the authors establish time-resolved high harmonic generation (tr-HHG) as a powerful spectroscopy method for tracking photoinduced dynamics in strongly correlated materials through a detailed investigation of the insulator-to-metal phase transitions in vanadium dioxide. Further examples include the mentioned studies of detection of Majorana fermions \cite{BBG22,pattanayak21}. 

Very recently, MBI group of Smirnova and Ivanov, using advanced techniques of dynamical mean field theory to study the strong field response of  the two-band Hubbard model (see M. Ivanov's contribution to \cite{ATTOVIII,VGE22} proposed sub-cycle multidimensional spectroscopy of strongly correlated materials).  In this work they introduce a new type of non-linear approach that allows  to unravel the sub-cycle dynamics of strongly correlated systems interacting with few-cycle infrared pulses For the two-dimensional Hubbard model under the influence of ultra-short, intense electric field transients, they demonstrate that our approach can resolve pathways of charge and energy flow between localized and delocalized many-body states on the sub-cycle timescale, including the creation of a highly correlated state surviving after the end of the laser pulse. 

HHG may be also used for extraction of higher-order nonlinear electronic response in solids \cite{HOK19}. 
Moreover, high harmonics may undergo spectacular enhancement in solid state nano-structures, such as
graphene heterostructures \cite{CRT22}. Very recently, HHG was used to characterize quantum criticality in strongly correlated systems \cite{SLZ22}. By employing the exact diagonalization method, we investigate the high-harmonic generation (HHG) of the correlated systems under the strong laser irradiation. For the extended Hubbard model on a periodic chain, HHG close to the quantum critical point (QCP) is more significant compared to two neighboring gapped phases (i.e., charge-density-wave and spin-density wave states), especially in low frequencies.

\begin{figure}[h!]
    \centering
    \includegraphics[width=\textwidth]{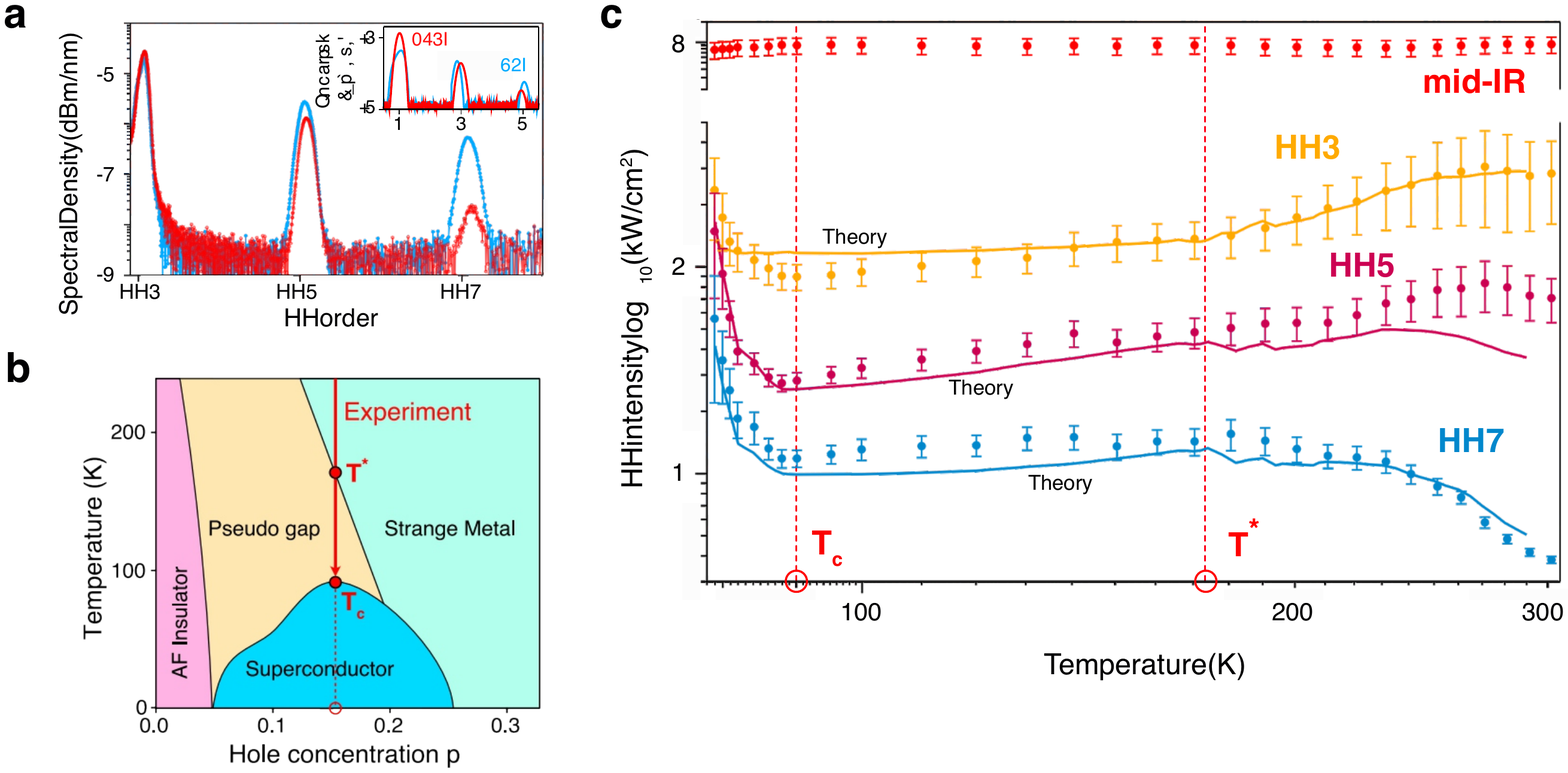}
    \caption{(a) Harmonic spectra showing odd orders HH3, HH5 and HH7 for room temperature (red) and at $T_\text{c} = 90 K$ (blue) for a mid-IR field strength of 0.083 V/{\AA}. Visible is already a blueshift of room-temperature harmonics with increasing harmonic order and relative to the harmonics measured at $T_\text{c}$. We observe a dramatic increase of HH7 amplitude upon cooling into the SC phase. The inlay shows the theoretical prediction of the spectrum, and the relative heights of the peaks matches well the behavior of the experimental data. (b) Shown is the phase diagram for YBCO. Based on the value of $T_\text{c}$ from the measurement, we determine a hole concentration of $p \sim 0.15$. Based on this value, we mark the transition between the different phases by the red arrow. In accord with measurement and theory, $T^\text{*}$ marks the transition between strange metal and pseudogap phases at 173 K.  (c) Shown is the reflected mid-IR field together with harmonics HH3, HH5 and HH7 as a function of temperature. These measurements are taken for a mid-IR field strength of 0.083 V/{\AA}. Results from the strong-field quasi-Hubbard model are overlaid as solid lines. We observe a dramatic increase of HH7 amplitude upon cooling into the SC phase at 88 K. All harmonic orders show a clear turning point at the critical temperature $T_\text{c}$ and an exponential increase in amplitude. More subtle, but still clearly discernible is another critical point $T^\text{*}$ at 173 K, marking the transition from strange metal to the pseudogap phase. The functional behaviour is reproduced by the model. Figure adapted from~\cite{ABB22}.
    }
    \label{Fig:YBCO}
\end{figure}

Another exciting prospect for HHG is to probe quantum phases in strongly correlated high-temperature superconductors. In Ref.~\cite{ABB22}, we have investigated the cuprate YBa$_2$Cu$_3$O$_7$ (YBCO) in a broad temperature range from 80K to 300 K. The investigation probes different phases of the high-$T_c$ material (Fig.~\ref{Fig:YBCO}(b)). This work demonstrates that strong-field emission is able to distinguish between various strongly correlated phases: The superconducting phase is marked by a strong increase of emission at all odd harmonic frequencies (3rd, 5th, and 7th), whereas the transition from the pseudogap phase to the strange metallic phase is reflected by a intensity drop only in the highest harmonics (7th) (Fig.~\ref{Fig:YBCO}(a)). These experimental findings are reproduced by simulating the strong-field dynamics of a two-band model in the BCS mean-field regime and in the presence of phenomenological scattering terms (Fig.~\ref{Fig:YBCO}(c)).

All of there findings, discussed in this subsection, open a way to a regime of imaging and manipulating strongly correlated materials at optical rates, far beyond the multi-cycle approach employed in Floquet engineering of quantum systems. 
   
\subsection{Detection of chirality with HHG}   

Detection of chirality is somewhat similar to the detection of topological order with HHG, but mostly refers to molecular targets. Chirality is the norm among biological molecules and the technological importance of distinguishing molecules which differ only by their chirality cannot be overstated. Absolutely seminal results along this direction in 
the context of ultrafast imaging 
have been achieved by the group of O. Smirnova and collaborators \cite{CBP15} (see \cite{AOS22} for a recent review). In particular, in Refs. \cite{AN19,AO21} they introduce the concept of \emph{locally chiral light} -- light which is chiral within the electric-dipole approximation by means of its 3D chiral Lissajous figure -- and show how this type of chiral light can be synthesized and exploited to dramatically enhance the contrast between HHG spectra in molecules with opposite chiralities. They show that such contrast can be enhanced either at the level of angle-integrated harmonic emission or at the level of direction of harmonic emission, depending on the pattern of the light's local chirality over the interaction region. In Ref. \cite{AOI21}, they address the role of the transverse spin of the light (which emerges whenever light is confined to a small region of space) in ultra-fast chiral imaging. They show that the interaction of chiral molecules with a strong, tightly focused, few-cycle pulse results in harmonics with a polarization that depends on the molecular chirality.

In turn, we have recently studied the role of the orbital angular momentum (OAM) of photoelectrons in ultrafast chiral imaging \cite{POL22}. We found that chiral molecules subject to strong-field ionization with a few-cycle, IR, linearly polarized pulse serve as natural sources of \emph{twisted} photoelectrons
, with the twist depending on the molecular chirality. Besides exposing an alternative road towards ultrafast chiral imaging, this effects suggests intriguing perspectives regarding the role of the electron OAM in recollision-based phenomena in chiral molecules.

\section{Generation of topology, strongly correlated systems, chirality, etc.}

While AP and HHG have proven to be great detection tools,  ultrafast laser pulse and attophysics methods may also be used for generation of topological order, strongly correlated states, chirality, etc. These ideas are, of course, very close to the attempts to generate room temperature high $T_c$ superconductivity \cite{CavalieriHighT_c}, or Chern insulators \cite{MSS20}, which belong more to the domain of ultrafast, but Terahertz physics. 

In fact, we have already some experience with laser induced phase transitions; we have worked with the experimental group of S. Wall on Terahertz field control of in-plane orbital order in La$_{0.5}$Sr$_{1.5}$MnO$_4$ \cite{MCT15}. We have also started to work on laser induce fluctuating bonds superconductivity \cite{JDC20} in the hole-doped cuprates. But, for this section perhaps the most relevant is our recent Phys. Rev. B Lett. \cite{BCG22}, where we demonstrate the theoretical  possibility of generation of fermionic Chern insulator from twisted light with linear polarization in graphene-like material. 

Particularly interesting line of research connects generation of topology, strongly correlated systems, chirality, etc. with the use of more complex {\it structured laser light}, combining polarization and OAM effects. A perfect example is the light, which knots fractional-order knots in the dynamics of the electric field vector employing  the polarization state of light and superposition of the fundamental and doubled frequency \cite{PJV19}. Application of strong laser pulsed of this form to atomic targets lead to ``exotic'' conservation of torus-knot angular momentum in high-order harmonic generation \cite{PRS19}.  Combining two delayed circularly polarized pulses of frequency $\omega$ and $2\omega$, one can generate a light with a self-torque: extreme-ultraviolet beams of HHG with time-varying orbital angular momentum \cite{RDB19}.\\

As shown in Ref. \cite{MPM22}, \emph{locally chiral light} \cite{AN19,AO21}, another type of structured light, can be used to efficiently (i.e. at the level of the electric-dipole approximation) imprint 3D chirality on achiral matter, such as atoms. Locally chiral light sculpts a chiral orbital out of the initially achiral ground state. That is, it excites the electron into a superposition of excited states such that the resulting orbital acquires a chiral shape. 

\section{Generation of massively quantum correlated states.}

The generation of massively quantum correlated states is another spectacular line of research of the ATTOQUIS platform. This direction it is driven by the first and the very encouraging experimental and theoretical results published/submitted on this subject within the last year \cite{RLP22,SRL22,LCP21,SRM22,LLH21,Sta22,RSM22}. In the following sub-sections we emphasize on the two main ideas underlying this research direction. The "QED and conditioning in strongly laser driven interactions" (sub--section 4.1) and the "QED of strong field processes driven by non-classical light fields" (sub--section 4.2). 

\subsection{QED and conditioning in strongly laser driven interactions}

\begin{figure}[t]
    \centering
    \includegraphics[width=.9\textwidth]{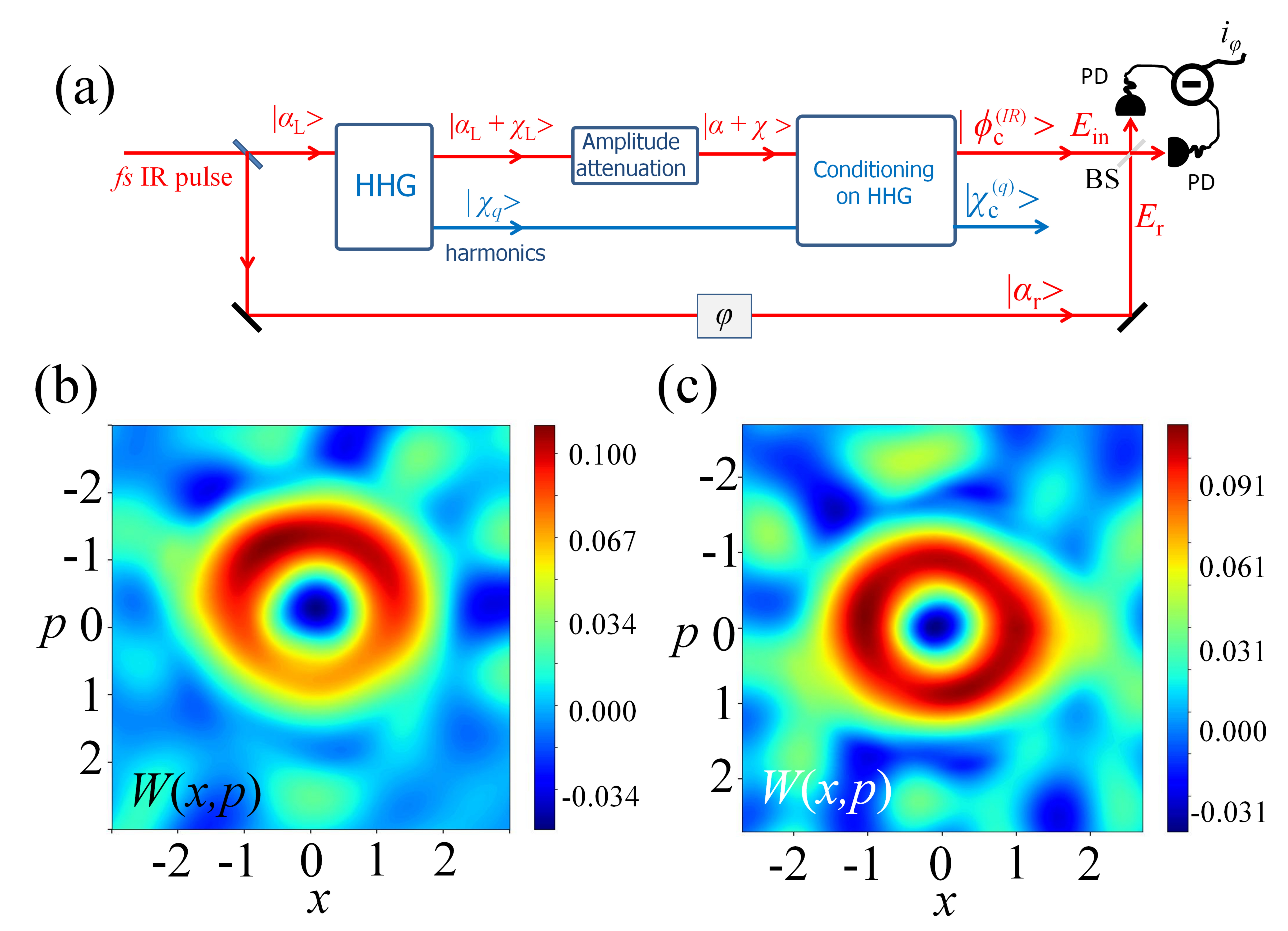}
    \caption{ (a) Schematic illustration of the operation principle of the conditioning approach. HHG is the area where the intense laser--matter interaction takes place and the high harmonics are generated. $|\alpha_{L}\rangle$ is the initial coherent state of the driving field. $|\alpha_{L}+\chi_{L}\rangle$, $|\alpha + \chi\rangle$ are the states of the IR field after the interaction and attenuation, respectively. $|\chi_{q}\rangle$ are the coherent states of the harmonic modes. $|\phi_{c}^{(IR)}\rangle$ and $|\chi_{c}^{(q)}\rangle$ are the field states after conditioning on the HHG. A scheme of the QT method used for the quantum state characterization. $|\alpha_r\rangle$ is the state of the local oscillator reference field. $\varphi$ is a controllable phase shift. BS is a beam splitter. PD are identical IR photodiodes used from the balanced detector. $i_{\varphi}$ is the $\varphi$ dependent output photocurrent difference used for the measurement of the electric field operator and the reconstruction of the Wigner function. (b) and (c) Reconstructed Wigner functions of an optical "cat" and "kitten" state, repsectively. Panels (b) and (c) have been reproduced from Ref.~\cite{RLP22}.}
    \label{Fig:conditioning}
\end{figure}

The basic idea of this approach, which is schematically illustrated in Fig.~\ref{Fig:conditioning}(a), is the following (here we focus in conditioning on HHG): 

\begin{itemize}

\item We describe the intense laser--matter interaction fully quantum-electrodynamically using coherent light states. In Fig.~\ref{Fig:conditioning} (a) the coherent state of the driving laser field is given by $|\alpha_{L}\rangle$.
\item During the interaction with the target, the driving pulse remains a coherent state, but is modified due to the interactions with the matter. The resulted driving field after the interaction is an amplitude shifted coherent state $|\alpha_{L}+\chi_{L}\rangle$ and the generated harmonics are in coherent state $|\chi_{q}\rangle$. These can be calculated using various methods. For atomic and molecular targets one can use TDSE or SFA \cite{ABC19}, for solids sometimes the semiconductor Bloch equations \cite{VMO14}, and for strongly correlated systems more sophisticated methods.
\item After passing the target, the fundamental laser field in attenuated in a coherent manner (resulting in $|\alpha + \chi\rangle$) in the range of few--photons and conditioned to HHG by means of quantum spectrometer (QS) method \cite{LCP21,SRM22,TKG17,TKD19}. The QS is a shot--to--shot IR vs HH photon correlation--based method, which selects only the IR shots that are relevant to the harmonic emission. It relies on photon statistics measurements and the energy conservation, i.e., $q$ IR photons need to be absorbed from the IR field for the generation of a photon of the $q_{th}$ harmonic. The output of the QS contains only the IR shots conditioned on the HHG process, resulting in the creation of non-classical light states. In case of intense laser--atom interaction, this state is a coherent state superposition i.e. an optical Schrödinger ``cat'' state of the form $|\phi_{c}^{(IR)}\rangle=|\alpha + \chi\rangle - \xi |\alpha\rangle$, where $\xi$ is the overlap between $|\alpha + \chi\rangle$ and $|\alpha\rangle$.

\item The quantum character of the light field after the conditioning ($E_{in}$) can be obtained by the measurement of its Wigner function. This can be archived by means of a homodyne detection system and the quantum tomography method \cite{LR09,BSM97}. The $E_{in}$ is spatiotemporally overlapped on a beam splitter (BS) with the field of a local oscillator $E_{r}$ (with $E_{r}>>E_{in}$). The $E_{r}$ is coming from the 2nd branch of the interferometer which introduces a controllable phase shift $\varphi$ between the $E_{in}$ and $E_{r}$. The fields after the BS are detected by a balanced differential photodetection system consisting of two identical photodiodes (PD). This provides at each value of $\varphi$ the photocurrent difference $i_{\varphi}$. The characterization of the quantum state of light can be achieved by recording for each shot the value of $i_{\varphi}$ as a function of $\varphi$. These values are directly proportional to the measurement of the electric field operator $\hat{E}_{in} (\varphi) \propto \cos(\varphi) \hat{x}+ \sin(\varphi) \hat{p}$ and are used for the reconstruction of the Wigner function via Radon transformation \cite{Herman80}. $\hat{x}=(\hat{a}+\hat{a}^\dagger)/\sqrt{2}$ and $\hat{p}=(\hat{a}-\hat{a}^\dagger)/i\sqrt{2}$ are the non-commuting quadrature field operators, and $\hat{a}$, $\hat{a}^\dagger$ are the photon annihilation and creation operators, respectively.
 
\item The scheme has been used for the generation of high--photon number shifted optical ``cat'' states (Fig.~\ref{Fig:Cats}) and coherent state superposition with controllable quantum features using intense--laser atom interaction \cite{LCP21,RLP22}. The control of the quantum features is shown in Figs.~\ref{Fig:conditioning}(b) and (c) where an optical ``cat'' state switches to a ``kitten'' for lower values of $\chi$. The same scheme can be applied for conditioning on ATI \cite{RLP22,SRM22,RSM22}, two-electron ionization, etc. In our recent study \cite{RLP22}, it has been theoretically shown that conditioning on ATI process can also lead to the generation of optical cat states. This, besides its fundamental interest associated with electron-photon correlation during the ATI/HHG process, provides an additional ``knob'' of controlling the quantum character of the optical cat states, a feature extremely valuable for applications in quantum technology. Also, according to our recent theoretical investigation performed for laser-atom interactions \cite{SRL22}, single and two-color driven laser-atom interactions can result in the generation of ``massively'' entangled optical coherent states in the spectral range from extreme ultraviolet (XUV) (shown in Fig.~\ref{Fig:conditioning}(a) as $|\chi_{c}^{(q)}\rangle$) to the far infrared. Such states in combination with passive linear optical elements (such as phase shifters, beam splitters, and optical fibers) \cite{RSP21,Sta22} can be considered as unique resources for novel applications in quantum technology. 
\end{itemize}

\subsection{QED of strong field processes driven by non-classical light fields}

All the quantum optical studies of strong field driven processes have thus far been approached by assuming a coherent state description for the driving field. The corresponding field is classical, even if Hilbert space methods are used for its description. 
However, in a recent work, presented at ATTO VIII by Even Tzur et al. \cite{Kaminar_squeezing}, this assumption was abandoned, and the process of HHG driven by non-classical light fields has been presented for the first time. 
They show, that the photon statistics of the driving field leaves its signatures in the observed high harmonic spectrum. 
This will open the path for investigating the interplay between non-classical properties of the driving source, and how they are imprinted in the observables of the harmonic field modes.

\section{Studies  of  quantum correlated states/decoherence in {\it Zerfall} processes}

\subsection{Generation of entanglement in {\it Zerfall} processes}

\begin{figure}
    \centering
    \includegraphics[width = 1.\textwidth]{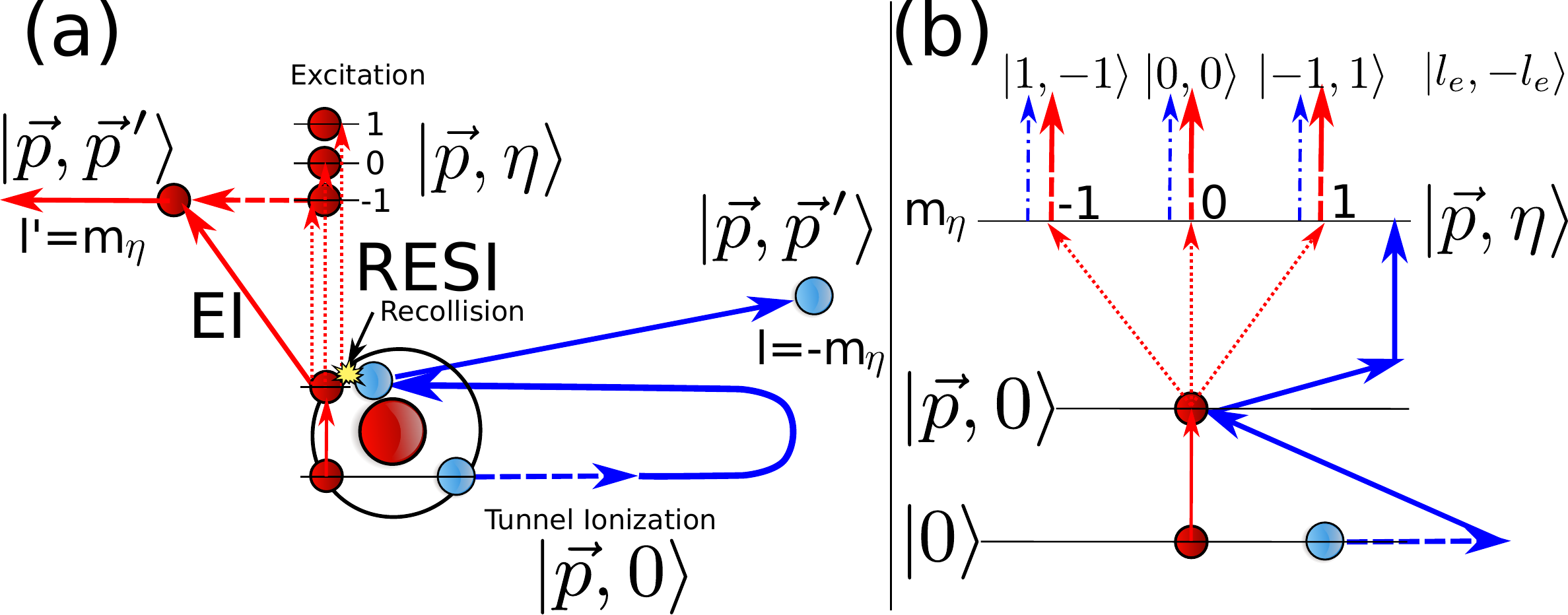}
    \caption{Non-sequential double ionization (NSDI), a `Zerfall' processes, leading to entanglement between the orbital angular momentum (OAM) of the two photoelectrons. In panel (a), an electron is ionized via a strong laser field and recollides with the parent atom/ molecule. This leads to either: direct ionization of a second electron in electron impact (EI) ionization, or excitation with subsequent ionization (RESI) of a second electron. Panel (b) focuses on RESI, the OAM superposition in the excited state is transferred to the final state along with anti-correlation between the electrons.}
    \label{Fig:NSDI}
\end{figure}

Finally, AP is a perfect playground to study generation of quantum correlations and entanglement in {\it Zerfall} processes, i.e. processes of decay of a “whole” in products. A prime example of this is non-sequential double ionization (NSDI), where strong-field ionization followed by laser induced recollision leads to double ionization of the target, see Fig.~\ref{Fig:NSDI}(a). Strong correlation between the ionization products, in particular the two photoelectrons, has long been known \cite{WGW00}. Previously, coherence and interference between the two photoelectrons has been studied in great detail \cite{MF15,MF16,HCL14,QHW17}. More recently, A. Maxwell {\it et al.} \cite{MML21} studied generation of entanglement in non-sequential double ionization, by using correlation in the OAM of the outgoing electrons, for a mini-review on OAM in strong-field ionization see \cite{MAC21,KPC21}. A key step, is to exploit the intermediate excited state, and its inherent superposition over OAM, which is present for the `second' electron for the low-intensity regime\footnote[1]{This is known as the recollision with subsequent ionization (RESI) mechanism.}. Thus, simple conservation laws dictate that the final OAM of the two electrons must be anti-correlated, which at certain final momentum leads to a maximally entangled qutrit, see Fig.~\ref{Fig:NSDI}(b), while remaining robust to incoherent effects, such as focal averaging or decoherence with the ion. 

Continuum products in a \textit{typical} strong-field Zerfall processes will be quantified through continuous variables, which can complicate analysis of entanglement measures. However, the use of OAM makes the task of quantifying and measuring entanglement much easier, the computations of the reduced density matrix of ionized product in the OAM space becomes simple, and in the case of NSDI, clearly exhibits entanglement. This also enables quantification by the, so called, logarithmic negativity, which may be used for mixed states to model incoherent effects. \\

\subsection{Characterizing decoherence in {\it Zerfall} processes}

Finally, studies of {\it Zerfall} processes offer a unique opportunity to understand and characterize quantum and classical decoherence. This goes back to seminal and pioneering theoretical works by 
R. Santra {\it et al.} \cite{PGH11,AVS17,ALK20},  F. Martín {\it et al.} \cite{LAT16}, or M. Vacher {\it et al.} \cite{VBR17}. All of these works dealt essentially with the single electron ionization process that might lead to entanglement of the electron with the parent ion. Measuring reduced density matrix of the electron provides thus information about decoherence and thus entanglement
Mark Vrakking {\it et al.} has developed both theoretical and the first experimental ideas in this context \cite{Vra21,KMD22} (see also M. Vrakking's contribution to \cite{ATTOVIII}). 
One of the main problems in this line of research is to make sure that the source of decoherence is truly quantum. Indeed, by analyzing with great care various classical and quantum models of decoherence in the process of single electron ionization, Ch. Bourassin-Bouchet {\it et al.} demonstrated dominantly classical sources of decoherence. 

In contrast, in the recent work \cite{BLF22}, A. L'Huillier {\it et al.} investigated
decoherence due to entanglement of the radial and angular degrees of freedom of the photoelectron.
They study two-photon ionization via the $2s2p$ autoionizing state in He using high spectral resolution
photoelectron interferometry. Combining experiment and theory, we show that the strong dipole
coupling of the $2s2p$ and $2p^2$ states results in the entanglement of the angular and radial degrees
of freedom. This translates, in angle integrated measurements, into a dynamic loss of coherence
during autoionization.

\section{Conclusions}
This paper provides a progress review on a new emerging interdisciplinary area of science, namely ATTOQUIS, in which ultrafast laser physics and attoscience merge with quantum optics and quantum information science. We provide examples of the ATTOQUIS research directions that are already at a step where theoretical and proof-of-principle experimental investigations have been developed. Based on these examples, we briefly discuss how the fully quantized description of intense laser-matter interaction and high--harmonic generation, in combination with conditioning methods, can be used for: i) the generation of controllable massively entangled quantum state superpositions in all states of matter, ii) the generation and detection of chirality and topological order in strongly correlated systems, iii) the generation of high--harmonics using intense quantum light, and iv) investigations quantum correlated states and decoherence effects in atoms. In few words, the results demonstrated in the aforementioned directions constitute the beginning of a very long "story". They show the power of intense laser--matter interaction as a new resource for novel investigations in attosecond science and quantum technology, a goal that the ATTOQUIS platform strives for. 

\section*{Acknowledgments}

M.L and group acknowledge support from ERC AdG NOQIA; Ministerio de Ciencia y Innovation Agencia Estatal de Investigaciones (PGC2018-097027-B-
\break
I00/10.13039/501100011033,  CEX2019-000910-S/10.13039/501100011033, Plan National FIDEUA PID2019-106901GB-I00, FPI, QUANTERA MAQS PCI2019-111828-2, QUANTERA DYNAMITE PCI2022-132919,  Proyectos de I+D+I “Retos Colaboración” QUSPIN RTC2019-007196-7); European Union NextGenerationEU (PRTR C17.I1);  Fundació Cellex; Fundació Mir-Puig; Generalitat de Catalunya (European Social Fund FEDER and CERCA program (AGAUR Grant No. 2017 SGR 134, QuantumCAT \ U16-011424, co-funded by ERDF Operational Program of Catalonia 2014-2020); Barcelona Supercomputing Center MareNostrum (FI-2022-1-0042); EU Horizon 2020 FET-OPEN OPTOlogic (Grant No 899794); National Science Centre, Poland (Symfonia Grant No. 2016/\break
20/W/ST4/00314); European Union’s Horizon 2020 research and innovation programme under the Marie-Skłodowska-Curie grant agreement No 101029393 (STREDCH) and No 847648  (“La Caixa” Junior Leaders fellowships ID100010434: LCF/BQ/PI19/11690013, LCF/BQ/PI20/11760031,  LCF/BQ/PR20/11770012, LCF/BQ/PR21/11840013); the Government of Spain (FIS2020-TRANQI and Severo Ochoa CEX2019-000910-S). M. F. C. acknowledges financial support from the Guangdong Province Science and Technology Major Project (Future functional materials under extreme conditions - 2021B0301030005).
J. B.and group acknowledge financial support from the European Research Council for ERC Advanced Grant “TRANSFORMER” (788218) and ERC Proof of Concept Grant “miniX” (840010), FET-OPEN “PETACom” (829153), FET-OPEN “OPTOlogic” (899794), EIC-2021-PATHFINDEROPEN "TwistedNano" (101046424), Laserlab-Europe (654148), Marie Sklodowska-Curie ITN “smart-X” (860553), Plan Nacional PID-PID2020-112664GB-I00-210901; National Science Centre, Poland (Symfonia Grant No. 2016/20/W/ST4/00314) AGAUR for 2017 SGR 1639, “Severo Ochoa” (SEV- 2015-0522), Fundació Cellex Barcelona, the CERCA Programme / Generalitat de Catalunya, and the Alexander von Humboldt Foundation for the Friedrich Wilhelm Bessel Prize.  P.~Tzallas group at FORTH acknowledges LASERLABEUROPE V (H2020-EU.1.4.1.2 grant no.
871124), FORTH Synergy Grant AgiIDA (grand no. 00133), the H2020 framework program for research and innovation
under the NEP-Europe-Pilot project (no. 101007417). ELI-ALPS is supported by the European Union and co--financed
by the European Regional Development Fund (GINOP Grant No. 2.3.6-15-2015-00001). J.R-D. acknowledges support from the Secretaria d'Universitats i Recerca del Departament d'Empresa i Coneixement de la Generalitat de Catalunya, as well as the European Social Fund (L'FSE inverteix en el teu futur)--FEDER. P. S. acknowledges funding from the European Union’s Horizon 2020 research and innovation programme under the Marie Sklodowska-Curie grant agreement No 847517. A. S. M. acknowledges funding support from the European Union’s Horizon 2020 research and innovation programme under the Marie Sk\l odowska-Curie grant agreement SSFI No.$\backslash 887153$.

%
%

\end{document}